\documentclass[a4paper,11pt]{article}
\pdfoutput=1
\usepackage{amsmath,eurosym,amssymb,amsfonts,amsfonts,wasysym,latexsym, multirow, multicol,bm,color}
\usepackage{graphicx}
\RequirePackage[numbers,sort&compress]{natbib}

\begin{document}

\renewcommand{\thefootnote}{\fnsymbol{footnote}}
\begin{center}
{\Large \bf Scaling attractors in interacting teleparallel dark energy}
\vskip 0.5cm
{\bf  G. Otalora}
\vskip 0.3cm
{\it Instituto de F\'{\i}sica Te\'orica, UNESP-Univ Estadual Paulista \\
Caixa Postal 70532-2, 01156-970 S\~ao Paulo, Brazil}

\vskip 0.8cm
\begin{quote}
{\bf Abstract.~}{\footnotesize It has been proposed recently the existence of a non-minimal coupling between a canonical scalar field (quintessence) and gravity in the framework of teleparallel gravity, motivated by similar constructions in the context of General Relativity. The dynamics of the model, known as teleparallel dark energy, has been further developed, but no scaling attractor has been found. 
Here we consider a model in which the non-minimal coupling is ruled by a dynamically changing coefficient $\alpha\equiv f_{,\phi}/\sqrt{f}$, with $f(\phi)$ an arbitrary function of the scalar field $\phi$. It is shown that in this case the existence of scaling attractors is possible, which means that the universe will eventually enter these scaling attractors, regardless of the initial conditions. As a consequence, the cosmological coincidence problem could be alleviated without fine-tunings. 
}
\end{quote}
\end{center}
\vskip 0.8cm

\section{Introduction}
Since the discovery of the universe accelerated expansion, this issue has been one of the most active fields in modern cosmology. The simplest explanation for the dark energy---the name given to the agent responsible for such accelerated expansion---is provided by a cosmological
constant. However, this scenario is plagued by a severe fine tuning problem associated with
its energy scale \cite{1,2,3}. This problem can be alleviated by considering a scalar field with a dynamically varying equation of state, like for example quintessence dark energy models. 
Scalar fields naturally arise in particle physics (including string theory), and these fields can act as candidates for dark energy. 
So far a wide variety of scalar-field dark energy models have been proposed, such as quintessence,  phantoms, k-essence, amongst many others \cite{1,4}. 

Usually, dark energy models are based on scalar fields minimally coupled to gravity, as well as a possible coupling of the field to a background fluid (dark matter). Also, a possible coupling between the scalar field and the Ricci scalar $R$ is not to be excluded in the context of generalized Einstein gravity theories. For example, a quintessence field coupled to gravity and called ``extended quintessence'' was proposed in Ref.~\cite{5} (see also Refs.~\cite{6,7}).
K-essence models non-minimally coupled to gravity were studied in Refs.~\cite{8,9}. It is also important to note that the first application of the non-minimal coupling to inflationary cosmology was done in Ref. \cite{Spokoiny}. 
Also, the application of non-minimally coupled models to quantum cosmology was considered in Ref. \cite{Barvinsky}. Finally, an important model of the
non-minimally coupled Higgs field has been considered in Ref. \cite{NonMinimallyHiggs}.

On the other hand,  Teleparallel Gravity or Teleparallel Equivalent of General Relativity \cite{GRwithtorsion,11,12,13} is an equivalent formulation of classical gravity, in which instead of using the torsionless Levi-Civita connection one uses the curvatureless Weitzenb\"{o}ck one. 
The dynamical objects are the four linearly independent tetrads, not the metric. The advantage of this framework is that the torsion tensor is formed solely from products of first derivatives of the tetrad.
Thus, apart from some conceptual differences, it is completely equivalent and indistinguishable from General Relativity (GR) at the level of field equations \cite{12,13}.
Recently the teleparallel gravity generalization, the so-called $F(T)$ theory \cite{14,15}, has attracted much attention.

In Refs.~\cite{16,17} it was proposed a non-minimal coupling between  quintessence and gravity in the framework of teleparallel gravity, motivated by a similar construction in the context of GR. The theory was called ``teleparallel dark energy'', and its dynamics was studied later in \cite{18,19}. However, no scaling attractor was found. 
Here we generalize the non-minimal coupling function $\phi^{2}\rightarrow f(\phi)$ and we define $\alpha\equiv f_{,\phi}/\sqrt{f}$ such that  $\alpha$ is constant for $f(\phi)\propto\phi^{2}$, but it can also be a dynamically changing quantity $\alpha(\phi)$. 
For dynamically changing $\alpha$ we show the existence of scaling attractors with accelerated expansion in agreement with observations. 
As long as the scaling solution is an attractor, for any generic initial conditions the system would sooner or later enter the scaling regime in which the dark energy density is comparable to the matter
energy density with accelerated expansion of the universe, alleviating in this the fine tuning problem of dark energy \cite{1,4}.

\section{Quintessence field in general relativity}

As is well known \cite{5,6,7}, one can generalize quintessence by including a non-minimal coupling between quintessence
and gravity. The relevant action reads
\begin{equation}
 S=\int d^{4}x\,\sqrt{-g}\,\left[\frac{R}{2\,\kappa^{2}}+\frac{1}{2}\,\partial_{\mu}\phi\,\partial^{\mu}\phi-V(\phi)+\xi\,\,f(\phi)\,R\right]+S_{m},
 \label{1}
\end{equation}where $R$ is the Ricci scalar, $S_{m}$ is the matter action, $\kappa^{2}=8\pi G$ and $c=1$ (we use the metric signature convention $(+,-,-,-)$ throughout). Also, the parameter $\xi$ is a dimensionless constant and $f(\phi)$ (with units of $mass^{2}$ and positive) is an arbitrary function of $\phi$  responsible for non-minimal coupling. 
In \cite{20,21,22,23} adopted $f\backsim \phi^{2}$ in the context inflation, other forms have been considered in \cite{24,25} (for a more general form that includes derivatives see \cite{26}).
When including a non-minimal coupling in quintessence, the corresponding Friedmann equations are preserved. On the other hand, the pressure density and energy density of quintessence  are changed to
\begin{multline}
 p_{\phi}=\frac{1}{2}\left(1+4\,\xi\,f_{,\phi\phi}\right)\dot{\phi}^{2}-V(\phi)+4\,\xi\left(f(\phi)+3\,\xi\,f_{,\phi}^{2}\right)\dot{H}- \\ 
 2\,\xi\,f_{,\phi}\,\dot{\phi}\,H-2\,\xi\,f_{,\phi}\,V_{,\phi}+6\,\xi\left(f(\phi)+4\,\xi\,f_{,\phi}^{2}\right)H^{2},
 \label{2}
\end{multline}
\begin{equation}
 \rho_{\phi}=\frac{1}{2}\,\dot{\phi}^{2}+V(\phi)-6\,\xi\,H\,f_{,\phi}\,\dot{\phi}-6\,\xi\,H^{2}\,f(\phi),
 \label{3}
\end{equation}where $V_{,\phi}\equiv dV/d\phi$, $f_{,\phi}\equiv df/d\phi$ and $H$ is the Hubble parameter. Also, a dot denotes differentiation with respect to the cosmic time $t$. For find $p_{\phi}$ and $\rho_{\phi}$ we have used the equation of motion
\begin{equation}
 \ddot{\phi}+3\,H\,\dot{\phi}-\xi\,R\,f_{,\phi}+V_{,\phi}=0,
 \label{4}
\end{equation}calculated from the action \eqref{1}, and the useful expression $R=6\,\left(\dot{H}+2\,H^{2}\right)$ in the flat Friedmann-Robertson-Walker (FRW) geometry, see \cite{6,7}.

\section{Teleparallel equivalent to general relativity}
Following \cite{13}, we briefly review the key ingredients of teleparallel gravity, a gauge theory for the translation group, which is fully equivalent to Einstein's general relativity. 
The tetrad field, the dynamical variable, forms an orthonormal basis for the tangent Minkowski space at each point $x^{\mu}$ of the spacetime and provides a relation between the tangent space metric $\eta_{a b}$ and the spacetime metric $g_{\mu\nu}$  through
\begin{equation}
 g_{\mu\nu}=h^{a}_{~\mu}\,h^{b}_{~\nu}\,\eta_{ab},
 \label{5}
\end{equation}
where $h^{a}_{~\mu}$ and the inverse $h_{a}^{~\mu}$ are the components of the tetrad field, which satisfy the orthogonality conditions $h^{a}_{~\mu}\,h_{a}^{~\nu}=\delta^{\nu}_{\mu}$ and $h^{a}_{~\mu}\,h_{b}^{~\mu}=\delta^{a}_{b}$. 
The tetrad field can be divided into a trivial part (non-gravitational) and a non-trivial part that is a translational gauge potential, which represents the gravitational field.

The action functional of teleparallel gravity can be written in the form
\begin{equation}
 S=\frac{1}{2\,\kappa^{2}}\int dx^{4}\,h\,T+S_{m},
 \label{6}
\end{equation}with $h=\det(h^{a}_{~\mu})=\sqrt{-g}$ and the torsion scalar $T$ is defined as
\begin{equation}
 T\equiv S_{\rho}^{~\mu\nu}\,T^{\rho}_{~\mu\nu},
 \label{7}
\end{equation}where
\begin{equation}
 T^{\rho}_{~\mu\nu}\equiv h_{a}^{~\rho}\,\left(\partial_{\mu}h^{a}_{~\nu}-\partial_{\nu}h^{a}_{~\mu}+A^{a}_{~e\nu}\,h^{e}_{~\mu}-A^{a}_{~e\mu}\,h^{e}_{~\nu}\right),
\label{8}
\end{equation}
is the torsion tensor and
\begin{equation}
 S_{\rho}^{~\mu\nu}\equiv \frac{1}{2}\left(K^{\mu\nu}_{~~\rho}+\delta^{\mu}_{\rho}\,T^{\theta\nu}_{~~\theta}-\delta^{\nu}_{\rho}\,T^{\theta\mu}_{~~\theta}\right),
 \label{10}
\end{equation}is the superpotential, with 
\begin{equation}
 K^{\mu\nu}_{~~\rho}\equiv -\frac{1}{2}\left(T^{\mu\nu}_{~~\rho}-T^{\nu\mu}_{~~\rho}-T_{\rho}^{~\mu\nu}\right),
 \label{9}
\end{equation}the contortion tensor.  The spin connection of teleparallel gravity $A^{a}_{~e\mu}$, which represents only inertial effects, is given by
\begin{equation}
 A^{a}_{~e\mu}=\Lambda^{a}_{~d}(x)\,\partial_{\mu}\Lambda_{e}^{~d}(x),
 \label{connect}
\end{equation}where $\Lambda^{a}_{~b}(x^{\mu})$ is a local (point-dependent) Lorentz transformation. For this connection the curvature vanishes identically
\begin{equation}
 R^{a}_{~b\nu\mu}=\partial_{\nu}A^{a}_{~b\mu}-\partial_{\mu}A^{a}_{~b\nu}+A^{a}_{~e\nu}\,A^{e}_{~b\mu}-A^{a}_{~e\mu}\,A^{e}_{~b\nu}=0,
 \label{curvature}
\end{equation}but the torsion \eqref{8} is non-vanishing. This connection can be considered a kind of ``dual'' of the GR connection, which is a connection with vanishing torsion, but non-vanishing curvature.
Also, it is important to note that curvature and torsion are properties of connections, and not of space-time. Many different connections, each one with different curvature and torsion, can be defined on the very same metric spacetime. The teleparallel connection and the GR connection are the only two choices respecting the correct number of degrees of freedom of gravitation.
The linear connection corresponding to the spin connection $A^{e}_{~b\nu}$ is
\begin{equation}
 \Gamma^{\rho}_{~\nu\mu}=h_{a}^{~\rho}\left(\partial_{\mu}h^{a}_{~\nu}+A^{a}_{~b\mu}h^{b}_{~\nu}\right),
 \label{linearconnect}
\end{equation}which is called  Weitzenb\"{o}ck connection. This connection is related to the Levi-Civita connection $\bar{\Gamma}^{\rho}_{~\mu\nu}$ of GR by
\begin{equation}
 \Gamma^{\rho}_{~\mu\nu}=\bar{\Gamma}^{\rho}_{~\mu\nu}+K^{\rho}_{~\mu\nu}. 
 \label{L.C.connect}
\end{equation}
The variation of the teleparallel action \eqref{6} with respect to the tetrad fields yields the teleparallel version of the gravitational field equation
\begin{equation}
 h^{-1}\,\partial_{\mu}\left(h\,h_{a}^{~\rho}\,S_{\rho}^{~\mu\nu}\right)-h_{a}^{~\lambda}\,T^{\rho}_{~\mu\lambda}\,S_{\rho}^{~\nu\mu}-\frac{1}{4}\,h_{a}^{~\nu}\,T+h_{c}^{~\rho}\,A^{c}_{~a\sigma}\,S_{\rho}^{~\nu\sigma}=\frac{\kappa^{2}}{2}\,h_{a}^{~\rho}\,\Theta_{\rho}^{~\nu},
 \label{11}
\end{equation}
where $\Theta_{\rho}^{~\nu}$ stands for the symmetric energy-momentum tensor. Using \eqref{L.C.connect}, through a lengthy but straightforward calculation, it can be shown that the gravitational field equation \eqref{11} is equivalent to the Einstein's field equation.

Now, like in special relativity, it is possible to define the preferred class of inertial frames, in the context of teleparallel gravity it is also possible to define a preferred class of frames: the class that reduces to the inertial class in the absence of gravitation. In this preferred class of frames, the spin connection of teleparallel gravity vanishes everywhere
\begin{equation}
 A^{a}_{~b\mu}=0,
 \label{prefeframes}
\end{equation}and the Weitzenb\"{o}ck connection assumes the form 
\begin{equation}
 \Gamma^{\rho}_{~\nu\mu}=h_{a}^{~\rho}\,\partial_{\mu}h^{a}_{~\nu}.
\end{equation}Of course, once a class of frames is specified, teleparallel gravity is no longer ``manifestly'' Lorentz invariant. For this reason, it is sometimes said that teleparallel gravity is not Lorentz invariant; that is, however, a wrong interpretation.
This would be equivalent to write down Maxwell theory in a specific gauge, and then to conclude that it is not gauge invariant. 
Anyway, in the same way that one can choose to work with electromagnetism in a specific gauge choice, one can also choose to work with teleparallel gravity in a specific class of frames \cite{13}.

From teleparallel gravity to cosmology, for a flat FRW background metric, the tetrad has the form
\begin{equation}
 h^{a}_{~\mu}(t)= \mbox{diag}(1,a(t),a(t),a(t)),
 \label{12}
\end{equation}where $a(t)$ is the scale factor. In the preferred class of inertial frames \eqref{prefeframes}, the tetrad choice \eqref{12} is an exact solution of the field equation \eqref{11}, and it is seen that the corresponding Friedmann
equations are identical to the GR ones both at the background and perturbation levels \cite{1,2,12,16}.  Also, when one introduces a scalar field as source of dark energy (for example quintessence field), in the case
of minimal coupling, there is no difference whether it is added in either of the two theories. However, this does not happen in the non-minimal case. In the non-minimal case the additional scalar sector is coupled to gravity, with
the torsion scalar in teleparallel gravity and with the curvature scalar in GR, and thus the resulting coupled equations do not coincide, which implies that the resulting theories are completely different.
This can be better understood if we take into account that the lagrangian of teleparallel gravity is equivalent to the lagrangian of general relativity  up to a divergence.

\section{Interacting teleparallel dark energy}
In what follows we study the action of teleparallel dark energy, in which a non-minimal coupling between an additional scalar sector and the torsion scalar of teleparallel gravity is considered. 
Also, we consider a possible interaction between teleparallel dark energy and dark matter, that is, ``interacting teleparallel dark energy'', as was studied in \cite{18}, since there is no physical argument to exclude the possible interaction between them.
We will work in the preferred class of frames \eqref{prefeframes}. The action will simply read
\begin{equation}
 S=\int d^{4}x\,h\,\left[\frac{T}{2\,\kappa^2}+\frac{1}{2}\,\partial_{\mu}\phi\,\partial^{\mu}\phi-V(\phi)+\xi\,f(\phi)\,T\right]+S_{m},
 \label{13}
\end{equation}
and the variation with respect to the tetrad field yields the coupled field equation 
\begin{multline}
2\,\left(\frac{1}{\kappa^{2}}+2\,\xi\,f(\phi)\right)\left[h^{-1}\,h^{a}_{~\alpha}\,\partial_{\sigma}\left(h\,h_{a}^{~\tau}\,S_{\tau}^{~\rho\sigma}\right)+T^{\tau}_{~\nu\alpha}\,S_{\tau}^{~\rho\nu}+\frac{T}{4}\,\delta^{~\rho}_{\alpha}\right]\\
+4\,\xi\,S_{\alpha}^{~\rho\sigma}\,f_{,\phi}\,\partial_{\sigma}\phi+\left[\frac{1}{2}\,\partial_{\mu}\phi\,\partial^{\mu}\phi-V(\phi)\right]\,\delta^{~\rho}_{\alpha}-\partial_{\alpha}\phi\,\partial^{\rho}\phi=\Theta_{\alpha}^{~\rho}.
 \label{14}
\end{multline}In flat FRW geometry \eqref{12}, the non-zero components of the torsion tensor, contortion tensor and the superpotential are
\begin{equation}
 T^{i}_{~0j}=-T^{i}_{~j0}=H\,\delta^{i}_{j},\:\:\:\:\: K^{0i}_{~~j}=-K^{i0}_{~~j}=H\,\delta^{i}_{j},\:\:\:\:\: S_{j}^{~i0}=-S_{j}^{~0i}=H\,\delta^{i}_{j},
\end{equation} where $i,j=1,2,3$. Therefore, imposing the flat FRW geometry \eqref{12} in \eqref{14}, we obtain the Friedmann equations  with
\begin{equation}
 \rho_{\phi}=\frac{1}{2}\,\dot{\phi}^{2}+V(\phi)-6\,\xi\,{H}^{2}\,f(\phi),
 \label{15}
\end{equation} the scalar energy density and
\begin{equation}
 p_{\phi}=\frac{1}{2}\,\dot{\phi}^{2}-V(\phi)+4\,\xi\,H\,f_{,\phi}\,\dot{\phi}+6\,\xi\,\left(3\,H^2+2\,\dot{H}\right)\,f(\phi),
 \label{16}
\end{equation}the pressure density of field. Here we also use the useful relation $T=-6\,H\,^{2}$, which arises for flat FRW geometry.

In this same background, the variation of the action \eqref{13} with respect to scalar field yields the field equation
\begin{equation}
 \ddot{\phi}+3\,H\,\dot{\phi}+V_{,\phi}+6\,\xi\,f_{,\phi}\,H^2=-\sigma.
 \label{17}
\end{equation}where $\sigma$ is the scalar charge corresponds to coupling between teleparallel dark energy and dark matter, and it is defined by the relation $\delta S_{m}/\delta\phi=-h\,\sigma$ \cite{1,4,31}.
Rewriting \eqref{17} in terms of $\rho_{\phi}$ and $p_{\phi}$  we find the continuity equation for the field
\begin{equation}
 \dot{\rho}_{\phi}+3\,H\,\rho_{\phi}\left(1+\omega_{\phi}\right)=-Q,
\label{18}
\end{equation}whereas for matter
\begin{equation}
  \dot{\rho}_{m}+3\,H\,\rho_{m}\left(1+\omega_{m}\right)=Q,
 \label{19}
\end{equation}where $\omega_{\phi}\equiv p_{\phi}/\rho_{\phi}$ and $\omega_{m}\equiv p_{m}/\rho_{m}=const\geq0$ are the equation-of-state parameter of teleparallel dark energy and dark matter, respectively.
The parameter $Q\equiv\dot{\phi}\,\sigma$ indicates the coupling between the two components.
Also,  we will define the barotropic index $\gamma\equiv1+\omega_{m}$ such that $1\leq\gamma<2$.

\section{Dynamical system of interacting teleparallel dark energy}

Of particular importance in the investigation of cosmological scenarios are those solutions in which the energy density of the scalar field mimics the background fluid energy density, $\rho_{\phi}/\rho_{m}=C$, with $C$ a nonzero constant. 
Cosmological solutions which satisfy this condition are called ``scaling solutions'' and the cosmological coincidence problem  can be alleviated in most dark energy models via 
scaling attractors (it is defined later) \cite{1,4}. To study the cosmological dynamics of the model, we introduce the followings dimensionless variables:
\begin{equation}
x\equiv\frac{\kappa\,\dot{\phi}}{\sqrt{6}\,H}\:\:\:\:\:\:\:\ y\equiv\frac{\kappa\,\sqrt{V}}{\sqrt{3}\,H}, \:\:\:\:\:\:\:\:\: u\equiv\kappa\,\sqrt{f},\:\:\:\:\:\:\:\:\:\:\: \lambda\equiv-\frac{V_{,\phi}}{\kappa\,V},\:\:\:\:\:\:\:\:\:\: \alpha\equiv\frac{f_{,\phi}}{\sqrt{f}}.
\label{20}
\end{equation}
Using these variables we define
\begin{equation}
s\equiv-\frac{\dot{H}}{H^{2}}=\frac{-3\,\gamma\,{y}^{2}+3\left( 2-\,\gamma\right) \,{x}^{2}+4\,\sqrt{6}\,\alpha\,\xi\,u\,x}{2\,\left( 2\,\xi\,{u}^{2}+1\right) }+\frac{3\,\gamma}{2},
\label{21}
\end{equation}
and the Eqs. \eqref{18} and \eqref{19} can be rewritten as a dynamical system of  ordinary differential equations (ODE), namely
\begin{equation}
 x'=\frac{\sqrt{6}}{2}\,\lambda\,{y}^{2}-\sqrt{6}\,\alpha\,\xi\,u+x\,\left(s-3\right)-\hat{Q},
 \label{22}
\end{equation}

\begin{equation}
 y'=\left( s-\frac{\sqrt{6}\,\lambda\,x}{2}\right) \,y,
 \label{23}
\end{equation}

\begin{equation}
 u'=\frac{\sqrt{6}\,\alpha\,x}{2},
 \label{24}
\end{equation}

\begin{equation}
 \lambda'=-\sqrt{6}\,\left( \Gamma-1\right) \,{\lambda}^{2}\,x,
 \label{25}
\end{equation}

\begin{equation}
 \alpha'=\sqrt{6}\,\left(\Pi-\frac{1}{2}\right)\,\frac{{\alpha}^{2}\,x}{u}.
 \label{26}
\end{equation}
In these equations primes denote derivative with respect to the so-called e-folding time $N\equiv\ln a$. Also, we
define
\begin{equation}
 \hat{Q}\equiv\frac{\kappa\,Q}{\sqrt{6}\,{H}^{2}\,\dot{\phi}}, \:\:\:\:\:\:\: \Pi\equiv\frac{f\,f_{,\phi\phi}}{f_{,\phi}^{2}},\:\:\:\:\:\:\:\:\:\: \Gamma\equiv\frac{V\,V_{,\phi\phi}}{V_{,\phi}^{2}}.
 \label{27}
\end{equation}

In terms of these dimensionless variables, the fractional energy densities $\Omega\equiv(\kappa^{2}\rho)/(3H^{2})$ for the scalar field and background matter are given by
\begin{equation}
 \Omega_{\phi}={y}^{2}+{x}^{2}-2\,\xi\,{u}^{2}, \:\:\:\:\:\:\:\: \Omega_{m}=1-\Omega_{\phi},
 \label{28}
\end{equation}respectively. 
Also, the equation of state of the field $\omega_{\phi}=p_{\phi}/\rho_{\phi}$ reads
\begin{equation}
 \omega_{\phi}=\frac{{x}^{2}-{y}^{2}+4\,\sqrt{\frac{2}{3}}\,\alpha\,\xi\,u\,x+2\left(1-\frac{2}{3}\,s\right)\,\xi\,{u}^{2}}{\left({y}^{2}+{x}^{2}\right)-2\,\xi\,{u}^{2}}.
 \label{29}
\end{equation}
On the other hand, the effective equation of state $\omega_{eff}=\left(p_{m}+p_{\phi}\right)/\left(\rho_{m}+\rho_{\phi}\right)$ is given by
\begin{equation}
\omega_{eff}=-\,\gamma\,{y}^{2}-\left(\gamma-2\right) \,{x}^{2}+4\,\sqrt{\frac{2}{3}}\,\alpha\,\xi\,u\,x+2\,\xi\left( \gamma-\frac{2}{3}\,s\right) \,{u}^{2}+\gamma-1,
\label{30}
\end{equation}and the accelerated expansion occurs for $\omega_{eff}<-1/3$. 

Here, we should emphasize that the dynamical system of ODE \eqref{22}-\eqref{26} are not a dynamical autonomous system unless the parameters $\Gamma$ and $\Pi$ are know. 
From now on we concentrate on exponential scalar-field potential of the form $V(\phi)=V_{0}\,e^{-\lambda\,\kappa\,\phi}$, such that $\lambda$ is a dimensionless constant, that is, $\Gamma=1$ (equivalently, we could consider potentials satisfying $\lambda\equiv-V_{,\phi}/\kappa\,V \thickapprox const$, which is valid for arbitrary but nearly flat potentials \cite{27,28,29}). In addition to the fact that exponential potentials can
give rise to an accelerated expansion (a power-law expansion $a(t)\propto t^{p}$ with $p>1$), they possess cosmological scaling solutions \cite{1,4}. 
On the other hand, for $f(\phi)\propto\phi^{2}$ or equivalently $\Pi=1/2$ then $\alpha\equiv f_{,\phi}/\sqrt{f}=const$. Also, for a general coupling function $u\equiv\kappa\,\sqrt{f(\phi)}$, with inverse function $\phi=f^{-1}(u^{2}/\kappa^{2})$, then $\alpha(\phi)$ and $\Pi(\phi)$  can be expressed in terms of $u$ (this approach is similar to that followed in the case of quintessence in GR with potential beyond exponential potential, see Ref. \cite{30}). 
Therefore, two situations may arise; one where $\alpha$ is a constant and  another where $\alpha$ depends on $u$. In both cases, we have a three-dimensional autonomous system \eqref{22}-\eqref{24}.

Once identified the autonomous system we can obtain the fixed points or critical points $(x_{c},y_{c},u_{c})$  by imposing the conditions $x'_{c}=y'_{c}=u'_{c}=0$. 
From the definition \eqref{20}, $x_{c}$, $y_{c}$, $u_{c}$  should be real, with $y_{c}\geq0$ and $u_{c}\geq0$.
To study the stability of the critical points, we substitute linear perturbations, $x\rightarrow x_{c}+\delta{x}$, $y\rightarrow y_{c}+\delta{y}$, and $u\rightarrow u_{c}+\delta{u}$  about the critical point $(x_{c},y_{c},u_{c})$ into the autonomous system \eqref{22}-\eqref{24} and linearize them. The eigenvalues of the perturbations matrix $\mathcal{M}$, namely, $\mu_{1}$, $\mu_{2}$ and $\mu_{3}$,  determine the conditions of stability of the critical points. One generally uses the following
classification \cite{1,4}: (i) Stable node: $\mu_{1}<0$, $\mu_{2}<0$ and $\mu_{3}<0$. (ii) Unstable node: $\mu_{1}>0$, $\mu_{2}>0$ and $\mu_{3}>0$. (iii) Saddle point: one or two of the three eigenvalues are positive and the other negative. (iv) Stable spiral: The determinant of the matrix
$\mathcal{M}$ is negative and the real parts of $\mu_{1}$, $\mu_{2}$ and $\mu_{3}$ are negative.  A  critical point is an attractor in the cases (i) and (iv), but it is not so in the cases (ii) and (iii). The universe will eventually enter these
attractor solutions regardless of the initial conditions.

Now we are going to study  the autonomous dynamical system \eqref{22}-\eqref{24}, first for  $\alpha=const\neq0$ and  then for  dynamically changing $\alpha(u)$, such that $\alpha(u)\rightarrow\alpha(u_{c})=0$  when the  system  falls into the critical point $(x_{c},y_{c},u_{c})$. 
Also, the fact that the energy density of teleparallel dark energy  $\Omega_{\phi}$ is of the same order as of dark matter $\Omega_{m}$ in the present universe, suggests that there may be some relation or interaction between them.
Several different forms of the coupling between dark energy and dark matter have been proposed. One possibility usually studied is to consider an interaction of the form $Q=\beta\,\kappa\,\rho_{m}\,\dot{\phi}$  with $\beta$ a dimensionless constant, see Refs. \cite{1,4,31}. 
A second approach is to introduce an interaction of the form $Q=\Upsilon\,\rho_{m}$ with the normalization of $\Upsilon$ in terms of the Hubble parameter $H$, i.e. $\Upsilon/H = \tau$, where $\tau$ is a dimensionless constant \cite{4}. Here we consider only the first possibility ($Q=\beta\,\kappa\,\rho_{m}\,\dot{\phi}$), for both, constant $\alpha$ and dynamically changing $\alpha$.

\section{Constant $\alpha$}

\subsection{Critical points}

In this section we consider a non-minimal coupling function $f(\phi)\propto\phi^{2}$ such that $\alpha=const\neq0$. From Eq. \eqref{27}, for the coupling  $Q=\beta\,\kappa\,\rho_{m}\,\dot{\phi}$ it is easy to find that
\begin{equation}
 \hat{Q}=\frac{\sqrt{6}\,\beta\,\Omega_{m} }{2},
 \label{32}
\end{equation}with $\Omega_{m}$ given in the equation \eqref{28}. Substituting \eqref{32} in \eqref{22} we can obtain the critical points of the autonomous system \eqref{22}-\eqref{24} for constant $\alpha$.
The critical points are presented in the table 1, and these are a generalization of the fixed points that were found in \cite{18,19}. In table 2 we summarize the stability properties (to be studied below), and conditions for acceleration and existence for each point.  The parameters $q_{\pm}$ are defined as
\begin{equation}
q_{\pm}=-\alpha\,\xi\pm\sqrt{\xi\,\left( {\alpha}^{2}\,\xi-2\,{\beta}^{2}\right) },
\label{33}
\end{equation}and the parameters $v_{\pm}$ as 
\begin{equation}
 v_{\pm}=\alpha\,\xi\pm\sqrt{\xi\,\left( {\alpha}^{2}\,\xi-2\,{\lambda}^{2}\right) }.
 \label{34}
\end{equation}
The point I.a is a matter-dominated solution ($\Omega_{m}=1$) with equation of state type cosmological constant $\omega_{\phi}=-1$, that exists for all values of $\xi$, $\lambda$ and $\beta=0$. 
Point I.b is a scaling solution and exists solely for $\xi<0$, $\alpha>0$ and $\beta>0$,  since  in this case the condition  $0\leq\Omega_{\phi}\leq1$ is satisfied.
The critical point I.c is also a scaling solution which there exists for $\xi<0$, but contrary to I.b, with $\alpha<0$ and $\beta<0$.
Accelerated expansion never occurs for these three points because $\omega_{eff}>-1/3$. Points I.d and I.e both correspond to dark-energy-dominated de Sitter solutions  with $\Omega_{\phi}=1$ and $\omega_{\phi}=\omega_{eff}=-1$. From \eqref{34}, the fixed point I.d exists for:
\begin{equation}
 \xi\geq 2\,{\lambda}^{2}/\alpha^{2}>0\:\:\:\: \text{and}\:\:\:\:\lambda/\alpha>0\:\:\:\: \text{or} \:\:\:\:\:\xi<0,\:\:\:\:\alpha<0\:\:\:\: \text{and} \:\:\:\:\lambda>0.
 \label{I.d}
\end{equation}

By the other hand, the point I.e exists for 
\begin{equation}
 \xi\geq 2\,{\lambda}^{2}/\alpha^{2}>0\:\:\:\: \text{and}\:\:\:\: \lambda/\alpha>0\:\:\:\: \text{or}\:\:\:\: \xi<0,\:\:\:\:\alpha>0\:\:\:\: \text{and}\:\:\:\:\lambda<0.
 \label{I.e}
\end{equation}
We note that the points I.d and I.e exist irrespective of the presence of the coupling $\beta$, since $\Omega_{\phi}=1$ in these cases.

\begin{table}[t]
 \centering
 \caption{Critical points for the autonomous system \eqref{22}-\eqref{24} for constant $\alpha\neq0$.}
\begin{center}
\begin{tabular}{c c c c c c c}\hline
Name & $x_{c}$ & $y_{c}$ & $u_{c}$  & $\Omega_{\phi}$ & $\omega_{\phi}$ &$\omega_{eff}$\\\hline
I.a & $0$ & $0$ & $0$ &  $0$ & $-1$ & $\gamma-1$\\\hline
I.b & $0$ & $0$ & $\frac{q_{-}}{2\,\beta\,\xi}$ &  $-\frac{q_{-}^{2}}{2\,{\beta}^{2}\,\xi}$ & $\gamma-1$ & $\gamma-1$\\\hline
I.c & $0$ & $0$ & $\frac{q_{+}}{2\,\beta\,\xi}$ &  $-\frac{q_{+}^{2}}{2\,{\beta}^{2}\,\xi}$ & $\gamma-1$ & $\gamma-1$\\\hline
I.d & $0$ & $\sqrt{\frac{\alpha\,v_{-}}{\lambda^{2}}}$ & $\frac{v_{-}}{2\,\lambda\,\xi}$ & $1$ & $-1$ & $-1$\\\hline
I.e & $0$ & $\sqrt{\frac{\alpha\,v_{+}}{\lambda^{2}}}$ & $\frac{v_{+}}{2\,\lambda\,\xi}$ & $1$ & $-1$ & $-1$\\\hline
\end{tabular}
\end{center}
\end{table}

\begin{table}[t]
\caption{Stability properties, and conditions for acceleration and existence of the fixed points in table 1.}
 \centering
\begin{center}
\begin{tabular}{c c c c c c c}\hline
Name & Stability & Acceleration & Existence \\\hline
I.a &  Saddle  &     No        &   $\beta=0$  \\\hline
I.b &  Saddle  &     No        &   $\xi<0$, $\alpha>0$ and $\beta>0$  \\\hline
I.c &  Saddle  &    No     &  $\xi<0$, $\alpha<0$ and $\beta<0$ \\\hline
I.d &  Stable node or stable spiral, or saddle &  All values  & Eq. \eqref{I.d} \\\hline
I.e &  Stable node or stable spiral, or saddle   & All values   & Eq. \eqref{I.e} \\\hline
\end{tabular}
\end{center}
\end{table}

\subsection{Stability}

Substituting the  linear perturbations, $x\rightarrow x_{c}+\delta{x}$, $y\rightarrow y_{c}+\delta{y}$, and $u\rightarrow u_{c}+\delta{u}$  into the autonomous system \eqref{22}-\eqref{24} and linearize them, we find the follows components for the matrix of linear perturbations $\mathcal{M}$

\begin{equation}
\mathcal{M}_{11}=\frac{-3\,\gamma\,{y}^{2}_{c}+9\,\left(2- \gamma\right) \,{x}^{2}_{c}+8\,\sqrt{6}\,\alpha\,\xi\,u_{c}\,x_{c}}{2\,\left( 2\,\xi\,{u}^{2}_{c}+1\right) }-\frac{3\,\left(2-\gamma\right)}{2}-\frac{\partial\hat{Q}}{\partial x}|_{x_{c},y_{c},u_{c}},
\label{35}
\end{equation}

\begin{equation}
 \mathcal{M}_{12}=\sqrt{6}\,\lambda\,y_{c}-\frac{3\,\gamma\,x_{c}\,y_{c}}{2\,\xi\,{u}^{2}_{c}+1}-\frac{\partial\hat{Q}}{\partial y}|_{x_{c},y_{c},u_{c}},
 \label{36}
\end{equation}

\begin{equation}
 \mathcal{M}_{13}=\frac{6\,\xi\,u_{c}\,x_{c}\,\left(\gamma\,{y}^{2}_{c}-\left(2-\gamma\right)\,{x}^{2}_{c}\right) +4\,\sqrt{6}\,\alpha\,\xi\,{x}^{2}_{c}}{{\left( 2\,\xi\,{u}^{2}_{c}+1\right) }^{2}}-\frac{2\,\sqrt{6}\,\alpha\,\xi\,{x}^{2}_{c}}{2\,\xi\,{u}^{2}_{c}+1}-\sqrt{6}\,\alpha\,\xi-\frac{\partial\hat{Q}}{\partial u}|_{x_{c},y_{c},u_{c}},
 \label{37}
\end{equation}

\begin{equation}
 \mathcal{M}_{21}=\left( \frac{3\,\left( 2-\gamma\right) \,x_{c}+2\,\sqrt{6}\,\alpha\,\xi\,u_{c}}{2\,\xi\,{u}^{2}_{c}+1}-\frac{\sqrt{6}\,\lambda}{2}\right) \,y_{c},
 \label{38}
 \end{equation}

 \begin{equation}
\mathcal{M}_{22}=\frac{-9\,\gamma\,{y}^{2}_{c}+3\,\left( 2-\gamma\right) \,{x}^{2}_{c}+4\,\sqrt{6}\,\alpha\,\xi\,u_{c}\,x_{c}}{2\,\left( 2\,\xi\,{u}^{2}_{c}+1\right) }-\frac{\sqrt{6}\,\lambda\,x_{c}-3\,\gamma}{2},
\label{39}
 \end{equation}
 
 \begin{equation}
  \mathcal{M}_{23}=\frac{6\,\xi\,u_{c}\,y_{c}\,\left( \gamma\,{y}^{2}_{c}+\left(2- \gamma\right)\,{x}^{2}_{c}\right) +4\,\sqrt{6}\,\alpha\,\xi\,x_{c}\,y_{c}}{{\left( 2\,\xi\,{u}^{2}_{c}+1\right) }^{2}}-\frac{2\,\sqrt{6}\,\alpha\,\xi\,x_{c}\,y_{c}}{2\,\xi\,{u}^{2}_{c}+1},
 \label{40}
 \end{equation}
 
 \begin{equation}
 \mathcal{M}_{31}=\frac{\sqrt{6}\,\alpha}{2}, \:\:\:\:\:\: \mathcal{M}_{32}=0, \:\:\:\:\:\: \mathcal{M}_{33}=0.
 \label{41}
 \end{equation}
 
The eigenvalues of the matrix $\mathcal{M}$, for each critical point, are as follows:

\begin{itemize}
 \item Point I.a:
\end{itemize}

\begin{equation}
 \mu_{1,2}=\frac{3\,\left(2-\gamma\right)\,\left(-1\pm\sqrt{1-\frac{16\,{\alpha}^{2}\,\xi}{3\,\left(2-\gamma\right)^{2}}}\right)}{4},\:\:\:\:\:\:\:\: \mu_{3}=\frac{3\,\gamma}{2}.
 \label{42}
\end{equation}

\begin{itemize}
 \item Point I.b:
\end{itemize}

\begin{equation}
\mu_{1,2}=\frac{3\,\left(2-\gamma\right)\,\left(-1\pm\sqrt{1+\frac{16\,\alpha\,\sqrt{\xi\,\left( {\alpha}^{2}\,\xi-2\,{\beta}^{2}\right)}}{3\,{\left( \gamma-2\right)^{2}} }}\right)}{4}, \:\:\:\:\:\:\: \mu_{3}=\frac{3\,\gamma}{2}.
\label{43}
\end{equation}

\begin{itemize}
 \item Point I.c:
\end{itemize}

\begin{equation}
\mu_{1,2}=\frac{3\,\left(2-\gamma\right)\,\left(-1\pm\sqrt{1-\frac{16\,\alpha\,\sqrt{\xi\,\left( {\alpha}^{2}\,\xi-2\,{\beta}^{2}\right)}}{3\,{\left( \gamma-2\right) }^{2}}}\right)}{4}, \:\:\:\:\:\:\: \mu_{3}=\frac{3\,\gamma}{2}.
\label{44}
\end{equation}

\begin{itemize}
 \item Point I.d:
\end{itemize}

\begin{equation}
 \mu_{1,2}=\frac{3\,\left(-1\pm\sqrt{1-\frac{4}{3}\,\alpha\,\sqrt{\xi\,\left( {\alpha}^{2}\,\xi-2\,{\lambda}^{2}\right) }}\right)}{2}, \:\:\:\:\:\:\: \mu_{3}=-3\,\gamma.
 \label{45}
\end{equation}

\begin{itemize}
 \item Point I.e:
\end{itemize}

\begin{equation}
  \mu_{1,2}=\frac{3\,\left(-1\pm\sqrt{1+\frac{4}{3}\,\alpha\,\sqrt{\xi\,\left( {\alpha}^{2}\,\xi-2\,{\lambda}^{2}\right) }}\right)}{2}, \:\:\:\:\:\:\: \mu_{3}=-3\,\gamma.
  \label{46}
\end{equation}

The points I.a, I.b, and I.c are saddle points in any case, since $\mu_{2}<0$ and $\mu_{3}>0$. 
For $\xi<0$ and $\alpha<0$ or $\xi>2\,\lambda^{2}/\alpha^{2}$ and $\alpha<0$ point I.d is a saddle point. On the other hand, for 
\begin{equation}
 \frac{2\,\lambda^{2}}{\alpha^{2}}<\xi\leq\frac{\lambda^{2}}{\alpha^{2}}\,\left(1+\sqrt{1+\frac{9}{16\,\lambda^{4}}}\right),
 \label{47}
\end{equation}and $\alpha>0$ it is a stable node. Also, when $\xi>\frac{\lambda^{2}}{\alpha^{2}}\,\left(1+\sqrt{1+\frac{9}{16\,\lambda^{4}}}\right)$ then $\mu_{1}$ and $\mu_{2}$ are complex with real part negative and 
$\det(\mathcal{M})=-9\,\alpha\,\gamma\,\sqrt{\xi\,\left( {\alpha}^{2}\,\xi-2\,{\lambda}^{2}\right) }<0$  for $\alpha>0$. So, in this case point I.d is a stable spiral.

Finally, point I.e is a saddle point for $\xi<0$ and $\alpha>0$ or $\xi>2\,{\lambda}^{2}/\alpha^{2}$ and $\alpha>0$. On the other hand, for $\xi$ as in \eqref{47} and $\alpha<0$, it is a stable node.
When $\xi>\frac{\lambda^{2}}{\alpha^{2}}\,\left(1+\sqrt{1+\frac{9}{16\,\lambda^{4}}}\right)$ then $\mu_{1}$ and $\mu_{2}$ are complex with real part negative and  $\det(\mathcal{M})=9\,\alpha\,\gamma\,\sqrt{\xi\,\left( {\alpha}^{2}\,\xi-2\,{\lambda}^{2}\right) }<0$  for $\alpha<0$. In this case, point I.e is a stable spiral.

So, for constant $\alpha$, $\xi>2\,\lambda^{2}/\alpha^{2}$ and $\beta=0$, the universe enters the matter-dominated solution I.a with $\omega_{\phi}=-1$ and $\omega_{eff}=0$ for $\gamma=1$ (non-relativistic dark matter), but since this solution is unstable, finally the system falls into the attractor I.d or I.e (dark-energy-dominated de Sitter solution) with $\Omega_{\phi}=1$ and $\omega_{\phi}=\omega_{eff}=-1$. 
Moreover, for $\beta\neq0$ and $\xi<0$ the system enters in the unstable scaling solution I.b or I.c, but, in this case the de Sitter solutions I.d or I.e are not late-time attractors of the system.
Also, for constant $\alpha$ no scaling attractor was found, even when we allow the possible interaction between teleparallel dark energy and dark matter.

\section{Dynamically changing $\alpha$}
Now let us consider a general function of non-minimal coupling  $f(\phi)$ such that $\alpha$ depends on $u$ and $\alpha(u)\rightarrow\alpha(u_{c})=0$  when $(x,y,u)\rightarrow (x_{c},y_{c},u_{c})$   ($(x_{c},y_{c},u_{c})$ is a fixed point of the system).
The field $\phi$ rolls down toward $\pm \infty$ ($x>0$ or $x<0$) with $f(\phi)\rightarrow u_{c}^{2}/\kappa^{2}$  when $(x,y,u)\rightarrow (x_{c},y_{c},u_{c})$. In this section, for simplicity,  we shall consider the
situation with positive values of $\lambda$ and $\beta$.

\subsection{Critical points}

The  critical points of the autonomous system \eqref{22}-\eqref{24} (with $\alpha$ replaced by $\alpha(u)$) are presented in  table 3. In table 3 we define $\hat{\xi}\equiv 2\,\xi\,u_{c}^{2}+1$ and $\zeta\equiv\lambda+\beta>0$. 
In table 4 we summarize the stability properties, and conditions for acceleration and existence for each point.

Point II.a corresponds to a scaling solution. The condition $0\leq\Omega_{\phi}\leq1$ is satisfied for $\beta<\sqrt{3/8}\,\left(2-\gamma\right)$ if
\begin{equation}
\frac{3\,\left(2-\gamma\right)^{2}}{4\,\beta^{2}}\,\left(1+\sqrt{1-\frac{8\,\beta^{2}}{3\,\left(2-\gamma\right)^{2}}}\right)\leq\hat{\xi}\leq\frac{3\,\left(2-\gamma\right)^{2}}{2\,\beta^{2}},
\label{48}
\end{equation}or 
\begin{equation}
0\leq\hat{\xi}\leq\frac{3\,\left(2-\gamma\right)^{2}}{4\,\beta^{2}}\,\left(1-\sqrt{1-\frac{8\,\beta^{2}}{3\,\left(2-\gamma\right)^{2}}}\right)<\frac{3\,\left(2-\gamma\right)^{2}}{2\,\beta^{2}}.
 \label{49}
\end{equation}On the other hand, for $\beta\geq\sqrt{3/8}\,\left(2-\gamma\right)$ it is required merely that

\begin{equation}
0\leq\hat{\xi}\leq\frac{3\,\left(2-\gamma\right)^{2}}{2\,\beta^{2}}.
\label{50}
\end{equation}

The equation of state for this solution is given by 
\begin{equation}
\omega_{\phi}=\frac{2\,{\beta}^{2}\,\left(2-\gamma\right) \,\hat{\xi}}{\left(\hat{\xi}-1\right)\,\left( 2\,{\beta}^{2}\,\left(\hat{\xi}-1\right)-3\,\left( \gamma-4\right) \,\gamma+4\,\left( {\beta}^{2}-3\right) \right) +2\,{\beta}^{2}}+\gamma-1,
\label{51}
\end{equation}and for $\hat{\xi}\approx1$ ($\xi\approx0$), and $\beta\neq0$, we have that $\omega_{\phi}\approx1$ (stiff matter). 

The condition of accelerated expansion is given by
\begin{equation}
\hat{\xi}<-\frac{\left(2-\gamma\right)\left(3\,\gamma-2\right)}{2\,\beta^{2}}<0.
\label{52}
\end{equation}Therefore, the accelerated expansion no occurs in this solution since necessary $\hat{\xi}$ is positive to ensure $0\leq\Omega_{\phi}\leq1$.

Points II.b and II.c exists for $\hat{\xi}\geq0$. Both points are scalar-field dominant solutions ($\Omega_{\phi}=1$), but without accelerated expansion because $\omega_{eff}=1$. 

Point II.d is also a scaling solution. From the condition $0\leq\Omega_{\phi}\leq1$ we have the following constraint
\begin{equation}
\frac{3\,\gamma}{\lambda\,\zeta}\leq\hat{\xi}\leq\frac{3\,\gamma}{\lambda\,\zeta}+\frac{\zeta}{\lambda}.
 \label{53}
\end{equation} 
The equation of state of field for this solution is given by 
\begin{equation}
\omega_{\phi}=\frac{\beta\,\zeta\,\gamma}{\beta\,\lambda\,\left(\hat{\xi}-2\right)+{\lambda}^{2}\,\left(\hat{\xi}-1\right)-3\,\gamma-{\beta}^{2}}+\gamma-1,
\label{54}
\end{equation}and for $\hat{\xi}\approx1$, $\lambda<<\beta$ ($\beta>>1$) we have that $\omega_{\phi}\approx-1$ (De Sitter solution). 
The universe exhibits an accelerated expansion independently of $\hat{\xi}$ for 
\begin{equation}
 \lambda<\frac{2\,\beta}{3\,\left(\gamma-1\right)+1}.
 \label{55}
\end{equation}In the case of $\gamma=1$ the accelerated expansion occurs for $\lambda<2\,\beta$.

Point II.e is a scalar-field dominant solution ($\Omega_{\phi}=1$) that exists for
\begin{equation}
0\leq\hat{\xi}\leq\frac{6}{\lambda^{2}},
 \label{56}
\end{equation}and gives an accelerated expansion at late times for
\begin{equation}
 \hat{\xi}<\frac{2}{\lambda^{2}}.
\label{57}
\end{equation}Given that $\hat{\xi}\geq0$ is necessary for the existence of this solution, then in table 3 we have that $\omega_{\phi}=\omega_{eff}=\lambda^{2}\,\hat{\xi}/3-1\geq-1$.

\begin{table}[t]
 \centering
 \caption{Critical points of the autonomous system \eqref{22}-\eqref{24} for dynamically changing $\alpha(u)$ such that $\alpha(u)\rightarrow\alpha(u_{c})=0$ and $u_{c}\geq0$. We define $\hat{\xi}\equiv 2\,\xi\,u_{c}^{2}+1$ and $\zeta\equiv \lambda+\beta$. }
\begin{center}
\begin{tabular}{c c c c c c c}\hline
Name & $x_{c}$ & $y_{c}$ & $u_{c}$  & $\Omega_{\phi}$ & $\omega_{\phi}$ &$\omega_{eff}$\\\hline
II.a & $-\frac{\sqrt{6}\,\beta\,\hat{\xi}}{3\,\left(2-\gamma\right) }$ & $0$ &  $u_{c}$  &  $\frac{2\,{\beta}^{2}\,{\hat{\xi}}^{2}}{3\,{\left( \gamma-2\right) }^{2}}+1-\hat{\xi}$ & Eq. \eqref{51} & $\frac{2\,{\beta}^{2}\,\hat{\xi} }{3\,\left(2- \gamma\right) }+\gamma-1$\\\hline
II.b & $-\sqrt{\hat{\xi}}$ &  $0$ &  $u_{c}$   &   $1$ & $1$ & $1$\\\hline
II.c & $\sqrt{\hat{\xi}}$  & $0$  &   $u_{c}$    &   $1$ & $1$ &$1$\\\hline
II.d & $\frac{\sqrt{6}\,\gamma}{2\,\zeta}$  & $\frac{\sqrt{\beta\,\zeta \,\hat{\xi} +\frac{3}{2}\,\left( 2-\gamma\right) \,\gamma}}{\zeta}$ &    $u_{c}$   & $-\frac{\lambda\,\hat{\xi} }{\zeta}+\frac{3\,\gamma}{{\zeta }^{2}}+1$ & Eq. \eqref{54} & $\frac{\lambda\,\left(\gamma-1\right)-\beta}{\zeta}$\\\hline
II.e & $\frac{\lambda\,\hat{\xi}}{\sqrt{6}}$  & $\sqrt{\hat{\xi} \,\left(1-\frac{\lambda^{2}\,\hat{\xi}}{6}\right) }$ &   $u_{c}$     &   $1$ & $\frac{\lambda^{2}\,\hat{\xi}}{3}-1$ & $\frac{\lambda^{2}\,\hat{\xi}}{3}-1$\\\hline
\end{tabular}
\end{center}
\end{table}

\begin{table}[t]
\caption{Stability properties, and conditions for acceleration and existence of the fixed points in table 3. As in table 3 we define $\hat{\xi}\equiv 2\,\xi\,u_{c}^{2}+1$ and $\zeta\equiv \lambda+\beta$.}
 \centering
\begin{center}
\begin{tabular}{c c c c c c c}\hline
Name & Stability & Acceleration & Existence \\\hline
II.a &  Saddle  &     No        &  Eqs. \eqref{48}, \eqref{49} and \eqref{50}    \\\hline
II.b &  Unstable node or saddle   &    No     & $\hat{\xi}\geq0$\\\hline
II.c & Unstable node or saddle  &   No   & $\hat{\xi}\geq0$ \\\hline
II.d &  Stable node or stable spiral or saddle   &  $\lambda<\frac{2\,\beta}{3\,\left(\gamma-1\right)+1}$   & $\frac{3\,\gamma}{\lambda\,\zeta}\leq\hat{\xi}\leq\frac{3\,\gamma}{\lambda\,\zeta}+\frac{\zeta}{\lambda}$ \\\hline
II.e &  Stable node or saddle    &  $\hat{\xi}<\frac{2}{\lambda^{2}}$ & $0\leq\hat{\xi}\leq\frac{6}{\lambda^{2}}$ \\\hline
\end{tabular}
\end{center}
\end{table}

\subsection{Stability}

For dynamically changing $\alpha(u)$, such that $\alpha(u)\rightarrow\alpha(u_{c})=0$, the components of the matrix of perturbation $\mathcal{M}$ are written as

\begin{equation}
\mathcal{M}_{11}=\frac{3\,\left(3\,\left(2-\gamma\right) \,{x}^{2}_{c}-\gamma\,{y}^{2}_{c}\right)}{2\,\hat{\xi}}-\frac{3\left(2-\gamma\right)}{2}-\frac{\partial\hat{Q}}{\partial x}|_{x_{c},y_{c},u_{c}},
\label{58}
\end{equation}

\begin{equation}
 \mathcal{M}_{12}=\sqrt{6}\,\lambda\,y_{c}-\frac{3\,\gamma\,x_{c}\,y_{c}}{\hat{\xi}}-\frac{\partial\hat{Q}}{\partial y}|_{x_{c},y_{c},u_{c}},
 \label{59}
\end{equation}

\begin{equation}
  \mathcal{M}_{13}=\frac{6\,\xi\,u_{c}\,x_{c}\,\left( \gamma\,{y}^{2}_{c}-\left(2-\gamma\right)\,{x}^{2}_{c}\right)}{\hat{\xi}^{2}}+\frac{2\,\sqrt{6}\,\xi\,\eta_{c}\,u_{c}\,{x}^{2}_{c}}{\hat{\xi}}-\sqrt{6}\,\xi\,\eta_{c}\,u_{c}-\frac{\partial\hat{Q}}{\partial u}|_{x_{c},y_{c},u_{c}},
 \label{60}
\end{equation}

\begin{equation}
 \mathcal{M}_{21}=\frac{3\,\left( 2-\gamma\right) \,x_{c}\,y_{c}}{\hat{\xi}}-\frac{\sqrt{6}\,\lambda\,y_{c}}{2},
 \label{61}
 \end{equation}

 \begin{equation}
\mathcal{M}_{22}=\frac{3\,\left(\left( 2-\gamma\right) \,{x}^{2}_{c}-3\,\gamma\,{y}^{2}_{c}\right)}{2\,\hat{\xi} }-\frac{\sqrt{6}\,\lambda\,x_{c}-3\,\gamma}{2},
\label{62}
 \end{equation}
 
 \begin{equation}
  \mathcal{M}_{23}=\frac{6\,\xi\,u_{c}\,y_{c}\,\left( \gamma\,{y}^{2}_{c}-\left(2-\gamma\right)\,{x}^{2}_{c}\right)}{\hat{\xi}^{2}}+\frac{2\,\sqrt{6}\,\xi\,\eta_{c}\,u_{c}\,x_{c}\,y_{c}}{\hat{\xi}},
  \label{63}
 \end{equation}
 
 \begin{equation}
 \mathcal{M}_{31}=0, \:\:\:\:\:\: \mathcal{M}_{32}=0, \:\:\:\:\:\: \mathcal{M}_{33}=\frac{\sqrt{6}\,\eta_{c}\,x_{c}}{2}.
 \label{64}
 \end{equation} Here we define $\eta_{c}\equiv \frac{d\alpha(u)}{d u}|_{u=u_{c}}$.  

Using \eqref{32} in \eqref{58}-\eqref{60} we can calculate the components of $\mathcal{M}$ for each critical point and the corresponding eigenvalues:

\begin{itemize}
 \item Point II.a:
\end{itemize}
\begin{equation}
 \mu_{1}=\frac{\beta\,\zeta\,\hat{\xi}}{2-\gamma}+\frac{3\,\gamma}{2}, \:\:\:\:\:\:\:\: \mu_{2}=-\frac{\beta\,\eta_{c}\,\hat{\xi} }{2-\gamma}, \:\:\:\:\:\:\:\: \mu_{3}=\frac{{\beta}^{2}\,\hat{\xi} }{2-\gamma}-\frac{3\,\left( 2-\gamma\right) }{2}.
 \label{65}
\end{equation}

\begin{itemize}
 \item Point II.b:
\end{itemize}

\begin{equation}
 \mu_{1}=-\frac{\sqrt{6}\,\eta_{c}\,\sqrt{\hat{\xi}}}{2},\:\:\:\:\:\:\: \mu_{2}=\frac{\sqrt{6}\,\left(\sqrt{6}+\lambda\,\sqrt{\hat{\xi}}\right)}{2}, \:\:\:\:\:\:\:\:\: \mu_{3}=-\sqrt{6}\,\beta\,\sqrt{\hat{\xi}}+3\,\left(2-\gamma\right). 
 \label{66}
\end{equation}

\begin{itemize}
 \item Point II.c:
\end{itemize}

\begin{equation}
\mu_{1}=\frac{\sqrt{6}\,\eta_{c}\,\sqrt{\hat{\xi}}}{2}, \:\:\:\:\:\: \mu_{2}=\frac{\sqrt{6}\,\left(\sqrt{6}-\lambda\,\sqrt{\hat{\xi}}\right)}{2}, \:\:\:\:\:\:\: \mu_{3}=\sqrt{6}\,\beta\,\sqrt{\hat{\xi}}+3\,\left(2-\gamma\right).
\label{67}
\end{equation}

\begin{itemize}
 \item Point II.d ($\gamma=1$):
\end{itemize}

\begin{equation}
\mu_{1,2}=\frac{3\,\left(\lambda+2\,\beta\right) }{4\,\zeta}\,\left(-1\pm\sqrt{1-\frac{8\,\left( 2\,\beta\,\zeta\,\hat{\xi}+3\right) \,\left(\zeta\,\lambda\,\hat{\xi}-3\right) }{3\,\left(\lambda+2\,\beta\right)^{2}\,\hat{\xi}}}\right), \:\:\:\:\:\:\: \mu_{3}=\frac{3\,\eta_{c}}{2\,\zeta}.
\label{68}
\end{equation}

\begin{itemize}
 \item Point II.e:
\end{itemize}

\begin{equation}
\mu_{1}={\lambda}\,\zeta\,\hat{\xi}-3\,\gamma, \:\:\:\:\:\:\:\: \mu_{2}=\frac{{\lambda}^{2}\,\hat{\xi}-6}{2}, \:\:\:\:\:\:\:\: \mu_{3}=\frac{\eta_{c}\,\lambda\,\hat{\xi}}{2}.
\label{69}
\end{equation}

Since for the point II.a  we have $0\leq\hat{\xi}\leq3\,\left(2-\gamma\right)^{2}/2\,\beta^{2}$ (Eqs.~\eqref{48}, \eqref{49} and \eqref{50}), then it is always a saddle point. 
The points II.b and II.c are either an unstable node or a saddle point. The point II.b is an unstable node for $0<\hat{\xi}<3\,\left(2-\gamma\right)^{2}/2\,\beta^{2}$ and $\eta_{c}<0$, whereas it is a saddle point for $\eta_{c}>0$ or  $\hat{\xi}>3\,\left(2-\gamma\right)^{2}/2\,\beta^{2}$.
On the other hand, the point II.c  is a  unstable node if $0<\hat{\xi}<6/\lambda^{2}$ and $\eta_{c}>0$, whereas for $\eta_{c}<0$ or $\hat{\xi}>6/\lambda^{2}$ it is a saddle point.

To study the properties of stability of the point II.d we shall consider the case where the background fluid is non-relativistic dark matter ($\gamma=1$). So, we restrict ourselves to positive $\hat{\xi}$ as in constraint \eqref{53} with $\gamma=1$, and  $\lambda<2\,\beta$ or equivalently $\zeta<3\,\beta$.

The eigenvalues $\mu_{1}$ and $\mu_{2}$ are real and negatives if 
\begin{equation}
\frac{3}{\zeta\,\lambda}<\hat{\xi}\leq\Delta+\frac{3}{\zeta\,\lambda},
 \label{70}
\end{equation}where
\begin{equation}
\Delta=\frac{3\,\left( \sqrt{{\delta}^{2}+64\,\beta\,\zeta\,{\left( \lambda+2\,\beta\right) }^{2}}+\delta\right) }{32\,\beta\,{\zeta}^{2}\,\lambda}>0,
\label{71}
\end{equation} with $\delta=\left( -7\,{\zeta}^{2}-6\,\beta\,\zeta+{\beta}^{2}\right)$. Then, for $\eta_{c}<0$ and $\hat{\xi}$ satisfying \eqref{70} the point II.d is a stable node. For $\eta_{c}>0$ and $\hat{\xi}$ as in \eqref{70} it is saddle point. 
On the other hand, for 
\begin{equation}
 \Delta+\frac{3}{\zeta\,\lambda}<\hat{\xi}\leq\frac{3}{\lambda\,\zeta}+\frac{\zeta}{\lambda},
 \label{72}
\end{equation} $\mu_{1}$ and $\mu_{2}$ are complex with real part negative, and the determinant of the matrix $\mathcal{M}$ that is given by
\begin{equation}
 \det(\mathcal{M})=\frac{9\,\eta_{c}\,\left( 2\,\beta\,\lambda\,{\zeta}^{3}\,{\hat{\xi}}^{2}+3\,{\zeta}^{2}\,\left(\lambda-2\,\beta\right) \,\hat{\xi}-9\,ζ\right) }{4\,{\zeta}^{4}\,\hat{\xi}},
 \label{73}
\end{equation}is negative if in addition $\eta_{c}<0$  (also in this case $\mu_{3}<0$). Therefore, for $\hat{\xi}$ as in \eqref{72} and $\eta_{c}<0$, point II.d becomes a stable spiral.  The constraint \eqref{70} is  consistent with the physical condition \eqref{53} (it can be stronger or weaker). On the other hand, the constraint \eqref{72} is also consistent with \eqref{53} provided that $\Delta<\zeta/\lambda$. In any case, either is a stable node or a stable spiral, point II.d is a scaling attractor with accelerated expansion if in addition we have $\lambda<2\,\beta$.

Point II.e is a stable node if 
\begin{equation}
0<\hat{\xi}<\frac{3\,\gamma}{\lambda\,\zeta}<\frac{6}{{\lambda}^{2}},
\label{74}
\end{equation}and $\eta_{c}<0$. Since $1\leq\gamma<2$ then $\frac{3\,\gamma}{\lambda\,\zeta}<\frac{6}{{\lambda}^{2}}$. On other hand, it is a saddle point for
\begin{equation}
 \frac{3\,\gamma}{\lambda\,\zeta}<\hat{\xi}<\frac{6}{{\lambda}^{2}}. 
 \label{75}
\end{equation}or $\eta_{c}>0$.
The point II.e is also an interesting solution, it is a scalar-field dominant solution ($\Omega_{\phi}=1$) and can be attractor (Eq. \eqref{74}) with accelerated expansion  (Eq. \eqref{57}) and such that $\omega_{\phi}=\omega_{eff}\geq-1$.

\begin{figure}[ht]
\centering
\includegraphics[width=0.55\textwidth]{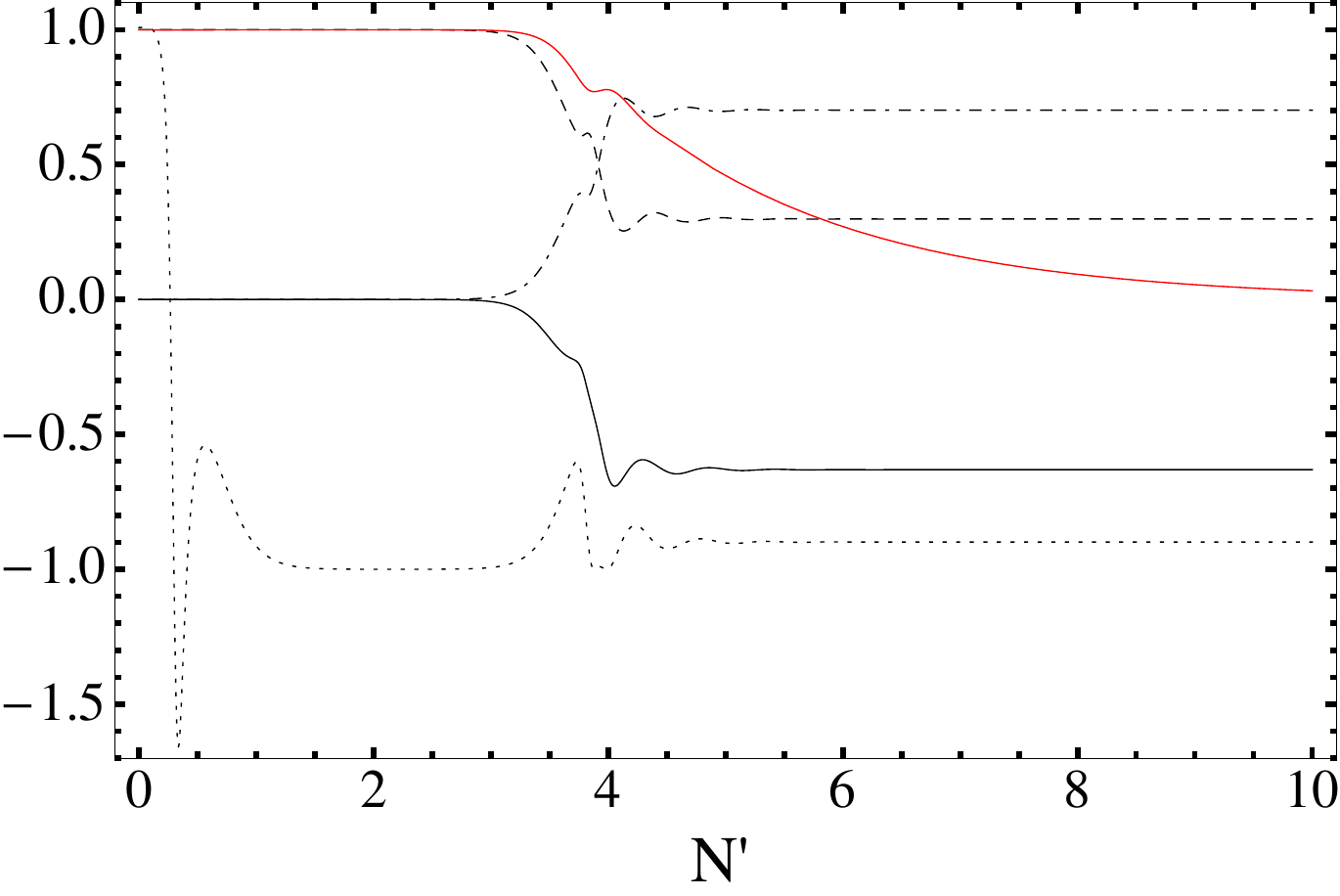}
\caption{Evolution of $\Omega_{m}$ (dashed), $\Omega_{\phi}$ (dotdashed), $\omega_{\phi}$ (dotted), $\omega_{eff}$ (solid) and $\alpha(u)$ (red line) with $\gamma=1$, $\lambda=2.4$ and $\xi\approx3.5\times10^{-5}$, in the case where the coupling to dark matter $\beta$ changes rapidly from $\beta_{1}=0$ to $\beta=4.1$ ($b=0.035$ and $u_{1}=0.21$). We choose initial
conditions $x_{i}=5\times10^{-5}$, $y_{i}=2.6\times10^{-6}$ and $u_{i}=7\times10^{-4}$ and by way of example we consider the function $\alpha(u)=u_{c}-u$ with $u_{c}=1$ and $\eta_{c}=-1$. The universe exits from the matter-dominated solution I.a with constant $\alpha$, $\omega_{\phi}=-1$, $\omega_{eff}=0$ and approaches the scaling attractor II.d for dynamically changing $\alpha(u)$ with $\Omega_{\phi}\approx0.70$, $\Omega_{m}\approx0.30$, $\omega_{\phi}\approx-0.90$ and $\omega_{eff}\approx-0.63$. We note that $\omega_{\phi}$ presents
the phantom-divide crossing during cosmological evolution.}
\label{figure}
\end{figure}

So, for dynamically changing $\alpha$ such that $\alpha(u)\rightarrow \alpha(u_{c})=0$, the fixed points II.d and II.e can attract the universe at late times and are equally viable solutions.
But since we are interested in the scaling attractors, now we are going to concentrate on the solution II.d. It is a scaling solution that has accelerated expansion and at the same time is an attractor (stable node or stable spiral).
For $\gamma=1$, the accelerated expansion occurs, independently of $\hat{\xi}$, for $\lambda<2\,\beta$ (Eq. \eqref{55}), and the physical condition $0\leq\Omega_{\phi}\leq1$ is ensures for $\hat{\xi}$ in accordance with \eqref{53}.
Also, for $\hat{\xi}$ in accordance with \eqref{70} and $\eta_{c}<0$ point II.d is a stable node, or for $\hat{\xi}$ as in \eqref{72} and $\eta_{c}<0$ it is a stable spiral.
When point II.d is a stable node, the constraint \eqref{70} can be stronger or weaker that the physical condition \eqref{53}.
Numerically we find that exist $\lambda_{*}(\beta)<2\,\beta$ such that for $\lambda<\lambda_{*}(\beta)$ then $\Delta>\zeta/\lambda$ and the constraint \eqref{70} is weaker that \eqref{53}. For example, for $\beta=0.6$ then $\lambda_{*}(\beta)\approx0.82$, and if $\lambda=0.2$ the constraint \eqref{70} implies $18.75<\hat{\xi}<22.96$. Moreover, the physical condition \eqref{53} implies $18.75<\hat{\xi}<22.75$. 
Also, in this example, for  $\hat{\xi}\approx19.95$ we have $\omega_{\phi}\approx-1.07$ (Eq. \eqref{54}), $\Omega_{\phi}\approx0.70$ and $\Omega_{m}\approx0.30$ (with accelerated expansion since $\lambda<2\,\beta$). However, this is not a suitable range for $\hat{\xi}$, since from solar system experiments (see for instance \cite{6}), necessary $\hat{\xi}\approx1$ (or $\xi\approx0$).
Finally, when point II.d is a stable spiral, that is, $\hat{\xi}$ in accordance with \eqref{72} and $\eta_{c}<0$, the physical condition \eqref{53} is ensures whenever $\Delta<\zeta/\lambda$. 
For example, for $\lambda=2.4$ and $\beta=4.1$ we have that \eqref{72} implies $0.23<\hat{\xi}<2.90$, whereas the constraint \eqref{53} implies $0.19<\hat{\xi}<2.90$. 
In Fig. 1 we show the case when the system approaches the fixed point II.d in the above example for $\lambda=2.4$, $\beta=4.1$ and for instance $\hat{\xi}\approx1.00064$ ($\xi\approx3.5\times10^{-5}$ and $u_{c}=1$).
In this figure, by way of example, we consider the function $\alpha(u)=u_{c}-u$ with $\eta_{c}=-1$. Also, following \cite{1, 32}, we consider a non-linear
coupling 
\begin{equation}
 \beta(u)=\frac{\left( \beta-\beta_{1}\right) \, \tanh\left( \frac{u-u_{1}}{b}\right) +\beta+\beta_{1}}{2},
 \label{76}
\end{equation} that changes between a small $\beta_{1}$ to a large $\beta$ in order to that initially the system enters in the matter-dominated solution I.a with $\beta_{1}=0$ and eventually approaches the scaling attractor II.d with $\omega_{\phi}\approx-0.90$, $\Omega_{\phi}\approx 0.70$, $\Omega_{m}\approx0.30$ and accelerated expansion ($\lambda<2\,\beta$).

\section{Conclusions}

It has been recently proposed \cite{16,17} a non-minimal coupling $\phi^{2}$  between quintessence and gravity in the framework of teleparallel gravity, motivated by a similar construction in the context of GR. The theory has been called ``teleparallel dark energy'', and the cosmological dynamics of the model was studied later in \cite{18,19}, but no scaling attractor was found. 
In this paper we generalize the function of non-minimal coupling  $\phi^{2}\rightarrow f(\phi)$ and we define $\alpha\equiv f_{,\phi}/\sqrt{f}$ such that  $\alpha$ is constant for $f(\phi)\propto\phi^{2}$, but it can also be a dynamically changing quantity $\alpha(\phi)$.
Also, in both cases, we consider a possible interaction between teleparallel dark energy and dark matter, that is, ``interacting teleparallel dark energy'',  as was studied in \cite{18}.

For constant $\alpha\neq0$, the universe enters the matter-dominated solution I.a ($\Omega_{m}=1$) with $\omega_{\phi}=-1$ and $\omega_{eff}=0$, but since this solution is unstable, it will eventually fall into the attractor I.d or I.e (dark-energy-dominated de Sitter solution) with $\Omega_{\phi}=1$ and $\omega_{\phi}=\omega_{eff}=-1$.
So, for constant $\alpha$ no scaling attractor was found, even when we allow interaction between teleparallel dark energy and dark matter (see table 1). 

On other hand, for dynamically changing $\alpha(\phi)$, and when it can be expressed in terms of $u\equiv\kappa\,\sqrt{f(\phi)}$ such that $\alpha(u)\rightarrow\alpha(u_{c})=0$ (we notice that $(x_{c},y_{c},u_{c})$ is a fixed point of the system), we show that interesting solutions exist (see table 3), in particular scaling attractors with accelerated expansion. These are non-minimal generalizations of the fixed points presented in \cite{1,31,33} for coupled dark energy in GR.
The final attractor is either the scalar-field dominated solution II.e or the scaling attractor II.d, both with accelerated expansion. The scaling solution II.d can be a stable node or a stable spiral when the background fluid is non-relativistic dark matter ($\gamma=1$).  When point II.d is a stable spiral, that is, $\hat{\xi}\equiv2\,\xi\,u_{c}^{2}+1$ as in constraint \eqref{72}, and 
\[
\eta_{c}\equiv \frac{d\alpha(u)}{du}|_{u=u_{c}}<0,
\]
it is possible to find a range for $\hat{\xi}$ in accordance with solar system experiments ($\hat{\xi}\approx1$) \cite{6}, and 
compatible with cosmological observations. For example, in Fig.~1 the system approaches point II.d with $\Omega_{\phi}\approx0.70$, $\Omega_{m}\approx0.30$, $\omega_{\phi}\approx-0.90$ and accelerated expansion, in agreement with observations. Also, we note that $\omega_{\phi}$ presents the phantom-divide crossing during cosmological evolution, as has been stressed in Refs.~\cite{16,17,19} for teleparallel dark energy.
Since the scaling solution II.d attracts the universe at late times regardless of the initial conditions with $\Omega_{\phi}$ comparable to $\Omega_{m}$ and accelerated expansion, the cosmological coincidence problem could be alleviated without fine-tunings \cite{1,4}.

As mentioned above, teleparallel dark energy, under the non-minimal coupling, is a completely different theory in relation to non-minimally coupled curvature gravity. 
The advantage of the former is that although the scalar field is canonical, one can obtain a dark energy sector being quintessence-like, phantom-like, or experiencing the phantom-divide crossing during the cosmological evolution, and so, the phantom regime can be described without the need of phantom fields,
which have ambiguous quantum behavior \cite{16}. A second important feature, which also happens in the scalar-field models with non-minimal derivative couplings, is that it does not exist conformal transformation that allows transit to an ``equivalent'', minimally-coupled theory with transformed field and potential, as it occurs in GR with a scalar field \cite{16,26}.
Also, given the rich structure of the theory, as in other models of dark energy with coupling to dark matter, in interacting teleparallel dark energy it is also possible to have scaling attractors with accelerated expansion.

\section*{Acknowledgments}
The author would like to thank J. G. Pereira and E.N. Saridakis for useful discussions and suggestions. He would like to thank also CAPES for financial support.

\end{document}